\documentclass[english,twocolumn]{article}
\usepackage[T1]{fontenc}
\usepackage[latin9]{inputenc}
\usepackage{color}
\usepackage{babel}
\usepackage{url}
\usepackage{amsmath}
\usepackage{amssymb}
\usepackage{graphicx}
\usepackage[]
 {hyperref}

\makeatletter
\newcommand{\lyxaddress}[1]{
	\par {\raggedright #1
	\vspace{1.4em}
	\noindent\par}
}

\makeatother

\begin{document}
\title{Utilizing Quantum Processor for the Analysis of Strongly Correlated
Materials}
\author{Hengyue Li$^{1}$\thanks{\protect\url{lihy@arclightquantum.com}},
Yusheng Yang$^{1}$ \thanks{\protect\url{yangys@arclightquantum.com}},
Pin Lv$^{2}$, Jinglong Qu$^{3,4,5}$, Zhe-Hui Wang$^{6}$, Jian Sun$^{1}$,
Shenggang Ying$^{1,7}$ \thanks{\protect\url{yingsg@ios.ac.cn}}}
\maketitle

\lyxaddress{1. Arclight Quantum Computing Inc., Beijing, 100191, China }

\lyxaddress{2. China Telecom Quantum Information Technology Group Co., Ltd.,
Hefei, 230031, China}

\lyxaddress{3. High-Temperature Materials Institute, Central Iron and Steel Research
Institute, Beijing, 100081, China}

\lyxaddress{4. Beijing GAONA Materials \& Technology Co., Ltd., Beijing, 100081,
China}

\lyxaddress{5. Sichuan CISRI-Gaona Forging Co., Ltd., Deyang, Sichuan, 618099,
China}

\lyxaddress{6. QuantumCTek (Shanghai) Co., Ltd., Shanghai, 200120, China}

\lyxaddress{7. Institute of Software, Chinese Academy of Sciences, Beijing, 100190,
China}
\begin{abstract}
This study introduces a systematic approach for analyzing strongly
correlated systems by adapting the conventional quantum cluster method
to a quantum circuit model. We have developed a more concise formula
for calculating the cluster's Green's function, requiring only real-number
computations on the quantum circuit instead of complex ones. This
approach is inherently more suited to quantum circuits, which primarily
yield statistical probabilities. As an illustrative example, we explored
the Hubbard model on a 2D lattice. The ground state was determined
utilizing \textit{Xiaohong}, a superconducting quantum processor equipped
with 66 qubits, supplied by \textit{QuantumCTek Co., Ltd}. Subsequently,
we employed the circuit model with controllable noise to compute the
real-time retarded Green's function for the cluster, which is then
used to determine the lattice Green's function. We conducted an examination
of the band structure in the insulator phase of the lattice system.
This preliminary investigation lays the groundwork for exploring a
wealth of innovative physics within the field of condensed matter
physics.
\end{abstract}

\section{Introduction}

Quantum computing garners interest for its potential to solve complex
problems beyond classical computing's reach. Pioneering algorithms
like Shor's algorithm \cite{Shor,Shor2} and Grover's algorithm \cite{grover}
have marked significant milestones. Additionally, emerging algorithms
such as the HHL algorithm \cite{HHL} are in development, expanding
the quantum computing repertoire. At the same time, quantum processor
hardware is also undergoing significant development. Unlike classical
computers, where manufacturing techniques have already converged on
electronic circuits based on silicon semiconductor materials, quantum
computing presents a unique challenge. Various technical approaches
are being explored for quantum computer realization, with superconducting
(SC) systems emerging as a frontrunner. Over the recent decades, there
has been notable progress not only in the number of qubits\textcolor{red}{{}
}but also in their quality in SC systems \cite{SC_review}. However,
achieving practical fault-tolerant quantum computing (FTQC) compatible
with standard quantum algorithms requires logical qubits of exceptionally
high quality. Among various error-correcting codes, the surface code
\cite{surface_code} has shown promise due to its high error threshold.
Although some demonstration experiments \cite{Krinner2022,PhysRevLett.129.030501}
have shown a modest improvement in the quality of qubits encoded using
surface code, these advancements are still insufficient for the implementation
of large-scale quantum computing systems.

A pragmatic strategy in the current stage of quantum computing is
the development of noisy intermediate-scale quantum (NISQ) applications.
Recent advancements in this area include the quantum variational method,
like variational quantum eigensolver (VQE) \cite{VQE_first,VQE_first_review}
and quantum machine learning algorithms \cite{Quantum_machine_learning,Supervised_Learning_with_Quantum_Computers}.
Representative algorithms of NISQ era, such as VQE, have found applications
in diverse fields, including quantum chemistry \cite{VQE_first,VQE_BeH2,VQE_H12}
and high-energy physics \cite{fit1}. It represents a hybrid quantum-classical
approach, utilizing quantum circuits to produce states that approximate
the ground state of a Hamiltonian, and employing a classical optimizer
to optimize across these circuits. However, applications of NISQ era
in other research fields, such as material science, are comparatively
less explored.

Materials exhibiting strong correlation effects, such as heavy-fermion
systems \cite{HeavyFermion} and non-Fermi-liquid systems \cite{NonFermi},
have long fascinated researchers. In recent decades, high-temperature
superconductors, perhaps one of the most attractive materials, have
also started to draw a lot of attention, promising to change the way
we use technology. Other interesting phenomena from strongly correlated
materials, like Mott metal-insulator transitions \cite{RevModPhys.40.677},
fractional quantum Hall effect \cite{RevModPhys.89.025005}, and spin-state
transitions \cite{spin1,spin2}, inspire the pursuit of new metamaterials.
All these materials don't behave as expected according to traditional
band theory \cite{bandtheory}. The excitement around these materials
comes from discovering unusual behaviors, such as novel quantum phase
transitions, and unexpected electric and magnetic properties. However,
simulating strongly correlated materials presents significant challenges
at the present time because of the complex interplay of electron-electron
interactions within these materials. Current simulation methods, predominantly
first-principles techniques like density functional theory (DFT),
face hurdles with systems containing strongly correlated electrons.
Concurrently, the development of non-perturbation theories has offered
alternative perspectives in quantum physics to simulate strongly correlated
materials. Dynamic mean field theory (DMFT) \cite{DMFT1,DMFT2} has
notably achieved early success by innovatively mapping the orbitals
in a lattice system to a local impurity model, based on the assumption
that the lattice self-energy is local. To fully consider the local
correlations, cluster perturbation theory (CPT) \cite{CPT} has subsequently
been developed. Additionally, considering broken-symmetry states,
other cluster methods with a meanfield approach have been developed
\cite{DCA1,CDMFT1,VCA0}. Collectively, these approaches for considering
local correlations are called quantum cluster methods (QCM) \cite{QCM1}. 

Leveraging NISQ circuits for quantum simulation of strongly correlated
systems appears appropriate and potentially sufficient \cite{Feynman1982}.
Various encoding methods \cite{JW,BK} enable the mapping of finite-sized
fermion clusters onto a circuit model. The core of QCM involves iteratively
calculating the Green's function of a cluster system. Several methods
\cite{Green1,Green2} for computing Green's functions on a circuit
model have been developed, making it theoretically possible to implement
QCM calculations on a quantum circuit model. It is the purpose of
this paper to introduce a systematic procedure for implementing each
critical step of QCM on the circuit model. As an illustrative example,
we investigate the paramagnetic insulator phase of a standard Hubbard
model on a 2-dimensional (2D) square lattice. The results obtained
solely through the circuit model will be compared with exact results
obtained from classical simulations. To account for the impact of
quantum gate imperfections, we introduce artificial noise into the
numerical simulation process.

In the remainder of this paper, we organize the content as follows.
Standard formulas and steps for QCM are revisited in Sec. \ref{QCM}.
The approach to determining the ground state via VQE is presented
in Sec. \ref{SecGS}. Section \ref{SecGF} introduces a streamlined
process for calculating the real-time retarded Green's function. Section
\ref{SecSteps} briefly summarizes the entire calculation procedure.
The results of the ground state energy, computed on the quantum processor
\textit{Xiaohong} provided by \textit{QuantumCTek}, are discussed
in Sec. \ref{SecGSE}. The findings regarding the cluster Green's
function are presented and analyzed in Sec. \ref{SecCGF}. Using QCM,
the one-particle excitation spectra are explored in Sec. \ref{SecAK}.
Finally, Sec. \ref{SecConc} offers a concise summary of our findings.

\section{\protect\label{QCM} Overview of the quantum cluster method}

We revisit fundamental theories of QCM for strongly correlated systems.
A general description of a fermion system can be expressed through
a Hamiltonian, as given by 

\begin{align}
H & =\sum_{ij}t_{ij}c_{i}^{\dagger}c_{j}+\sum_{ijkl}X_{ijkl}c_{i}^{\dagger}c_{j}^{\dagger}c_{k}c_{l},\label{eq:2.11}
\end{align}
where the operator $c_{i}^{\dagger}$ ($c_{j}$ ) creates (annihilates)
an electron at site $i$ ($j$). Without loss of generality, we suppress
the index of spin (and also other possible orbital degrees of freedom).
The hopping energy, denoted as $t_{ij}$, and the interacting energy,
$X_{ijkl}$ , as outlined in Eq. \ref{eq:2.11}, are typically computed
using \textit{ab initio} methods with proper basis sets. To illustrate,
the interacting energy $X_{ijkl}$ is defined by the integral 
\begin{align}
X_{ijkl} & =\int\int d\boldsymbol{r}d\boldsymbol{r}'\phi_{i}^{*}(\boldsymbol{r})\phi_{j}^{*}(\boldsymbol{r}')V(\boldsymbol{r}-\boldsymbol{r}')\phi_{k}(\boldsymbol{r}')\phi_{l}(\boldsymbol{r}),
\end{align}
where $\{\phi_{i}(\boldsymbol{r})\}$ represents a complete set of
states and $V(\boldsymbol{r}-\boldsymbol{r}')$ is the electron-electron
interacting energy \cite{Mahan}. This model is versatile and can
be adapted to represent many well-known models. For instance, by focusing
solely on the interacting term $X_{ijkl}\delta_{ik}\delta_{jl}$ and
assigning $i$ and $j$ to different spin indices on the same site,
Eq. \ref{eq:2.11} transforms into the standard Hubbard model \cite{Hubbard1},
described as: 
\begin{align}
H & =-\gamma\sum_{ij\sigma}\left(c_{i\sigma}^{\dagger}c_{j\sigma}+h.c.\right)+Un_{i\uparrow}n_{i\downarrow}-\mu\sum_{i\sigma}n_{i\sigma},\label{eq:22}
\end{align}
where $n_{i\sigma}=c_{i\sigma}^{\dagger}c_{i\sigma}$ is the particle
number for spin $\sigma$ on the $i$-th site, $\mu$ is the chemical
potential, $-\gamma$ is the hopping integral, and $U$ represents
the interaction term of electron repulsion. 

\begin{figure}
\includegraphics[bb=0bp 0bp 298bp 142bp]{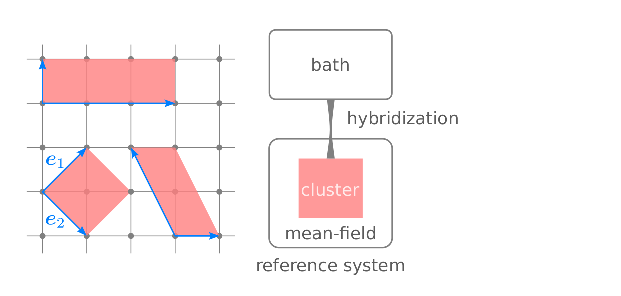}\caption{\protect\label{fig:qcm} The original lattice is divided into distinct
clusters and inter-cluster segments. Each cluster (red areas), referred
to as a reference system, is independently analyzed using various
methods, such as exact diagonalization (ED). The inter-cluster section
is treated as a perturbation affecting the reference system. To address
symmetry breaking, we introduce additional mean-fields. Meanwhile,
achieving the metallic phase requires integrating additional environmental
elements, typically bath sites, which are hybridized with the cluster.
These additional parameters, mean-fields, and hybridization functions
are determined through a self-consistent variational process.}
\end{figure}

In QCM, the original lattice is tiled through super-lattice expanded
by $\{\boldsymbol{e}_{i}\}$, with super-lattice-cell of size $L$,
as shown in Fig. \ref{fig:qcm}. For each operator $c_{i}$, a distinct
cluster index $\boldsymbol{R}$ and a corresponding in-site index
$\alpha$ are identified. This allows for the operator to be re-expressed
in a more refined form, transforming 

\begin{align}
c_{i} & \to\psi_{\boldsymbol{R}\alpha}.\label{eq:2.12}
\end{align}

With this notation, Eq. \ref{eq:2.11} is decoupled by tiling clusters
and can be rewritten as $H=\sum_{\boldsymbol{R}}H_{c}\left(\{\psi_{\boldsymbol{R}\alpha}\}\right)+\sum_{\boldsymbol{R}\boldsymbol{R}'}T(\boldsymbol{R}-\boldsymbol{R}')$.
Here, $H_{c}\left(\{\psi_{\boldsymbol{R}\alpha}\}\right)$ represents
a cluster-specific Hamiltonian, while $T(\boldsymbol{R}-\boldsymbol{R}')$
denotes the Hamiltonian terms that facilitate connections between
the cluster $\boldsymbol{R}$ and $\boldsymbol{R}'$. Each cluster
can be considered as a reference system for abstracting the self-energy
\cite{CPT}. The lattice Green's function, $\boldsymbol{\mathcal{G}}(\boldsymbol{q},\omega)$,
 is given by 
\begin{align}
\boldsymbol{\mathcal{G}}(\boldsymbol{q},\omega) & =\left[\boldsymbol{G}^{-1}(\omega)-\tau_{\boldsymbol{q}}\right]^{-1},\label{eq:3}
\end{align}
where $\boldsymbol{q}$ is the reciprocal vector in the reduced Brillouin
zone, $\boldsymbol{G}(\omega)$ (a $L\times L$ matrix) is the Green's
function of a local cluster, and $\tau_{\boldsymbol{q}}$ is a linear
combination of $T(\boldsymbol{0}-\boldsymbol{r})$ around a center
cluster $\boldsymbol{0}.$ Details of these derivations can be found
in Appendix \ref{Apd_qcm}. 

We recall some basic knowledge of Green's function theory for calculating
each matrix element $G_{ij}$ of $\boldsymbol{G}$. The definition
of the real-time retarded Green's function \cite{Economou2006-gd}
is given by 
\begin{align}
G_{ij}^{R}(t-t') & =-i\Theta(t-t')\langle\{c_{i}(t),c_{j}^{\dagger}(t')\}\rangle\label{eq:4}
\end{align}
where $\Theta$ is the Heaviside step function, the curly bracket
$\{*,*\}$ represents the anti-commutator, $c_{i}(t)=e^{iHt}c_{i}e^{-iHt}$
is the Heisenberg representation of the operator $c_{i}$, the bracket
$\langle...\rangle=\text{Tr}\left(e^{-H/T}...\right)/\text{Tr}\left(e^{-H/T}\right)$
represents the thermal average. Throughout this paper, we set $\hbar=1$.
In zero temperature case $(T=0)$, the calculation of the thermal
average value is specialized by $\langle g|...|g\rangle$, where $|g\rangle$
is the ground state of the system (assuming a non-degenerate scenario).
With relation $\Theta(t-t')=\frac{-1}{2\pi i}\int_{-\infty}^{\infty}d\omega\frac{e^{-i(t-t')\omega}}{\omega+i\eta},$
where $\eta\to0^{+}$ is an infinite small real number, we rewrite
the real-time retarded Green's function as 
\begin{align}
G_{ij}^{R}(t) & =\frac{1}{2\pi}\int_{-\infty}^{\infty}d\omega G_{ij}(\omega+i\eta)e^{-i\omega t},\label{eq:1}
\end{align}
where $G_{ij}(\omega+i\eta)$ is the Fourier transform of the real-time
retarded Green's function. With analytic continuation to full complex
plane $\omega+i\eta\to z$ and Kramers--Kronig relations \cite{KKrelation},
the Green's function is given by 
\begin{align}
G_{ij}(z) & =\int dz'\frac{\rho_{ij}(z')}{z-z'},
\end{align}
where 
\begin{align}
\rho_{ij}(\omega) & =-\frac{1}{\pi}\mathcal{I}\left[G_{ij}(\omega+i\eta)\right]\label{eq:2}
\end{align}
is the spectral function. Many useful Green's functions can be derived
from this form. For instance, the Matsubara Green's function can be
calculated by $G_{ij}(i\omega_{n})$, where $\omega_{n}$ is the Matsubara
frequency \cite{Mahan}. In Lehmann representation, at zero temperature
case the spectral function is given by 
\begin{align}
\rho_{ij}(\omega) & =\sum_{n}\langle g|c_{i}|n\rangle\langle n|c_{j}^{\dagger}|g\rangle\delta\left(\omega-E_{n}+E_{0}\right)\nonumber \\
 & +\sum_{n}\langle n|c_{i}|g\rangle\langle g|c_{j}^{\dagger}|n\rangle\delta\left(\omega-E_{0}+E_{n}\right),
\end{align}
where $\{|n\rangle\}$ and $\{E_{n}\}$ are eigenstates and eigenvalues
of the system respectively. When $i=j$, $\rho_{i}(\omega)\equiv\rho_{ij}(\omega)$
represents the density of states at orbital $i$. It obeys the summation
rule $\int\rho_{i}(\omega)=1$. Practically, we calculate the spectral
function by Eq. (\ref{eq:2}). And the Green's function in frequency
domain, $G_{ij}(\omega+i\eta),$ is calculated by the inverse of Eq.
(\ref{eq:1}) and given by 
\begin{align}
G_{ij}(\omega+i\eta) & =\int_{0}^{\infty}dte^{i\omega t}\left[e^{-\eta t}G_{ij}^{R}(t)\right].\label{eq:10}
\end{align}
In actual computations, we rely on an artificial broadening parameter
$\eta$ to facilitate the calculation, instead of employing an impractical
value of $0^{+}$. Henceforth, we disregard the subscript ``$R$''
in $G_{ij}^{R}(t)$.

\section{Calculation on circuits}

\subsection{\protect\label{SecGS} Ground state}

The QCM effectively segments the original lattice into distinct clusters.
These clusters, each characterized by a specific size labeled $L$,
are regarded as individual cluster systems. The VQE method is employed
to investigate the ground state of the system using quantum processor.
We employed a parameterized unitary transformation, denoted as $U_{G}(\boldsymbol{\theta})$,
to construct the parameterized state from scratch as $|g(\boldsymbol{\theta})\rangle=U_{G}(\boldsymbol{\theta})|\boldsymbol{0}\rangle.$
In VQE method, the expectation value of the Hamiltonian serves as
an upper limit for the ground state energy. Thus, the ground state
energy $E_{0}$ is bounded by $E_{0}\leq\langle g(\boldsymbol{\theta})|H_{c}|g(\boldsymbol{\theta})\rangle$,
where $H_{c}$ is the Hamiltonian of the system.

In order to performed VQE by quantum circuits, it is essential to
transform the Hamiltonian into a format compatible with quantum processor
measurements. We adopt the standard Jordan-Wigner (JW) encoding \cite{JW}:
\begin{align}
c_{i} & =\frac{1}{2}\left(X_{i}-iY_{i}\right)\prod_{\alpha<i}\left(-Z_{\alpha}\right),\label{eq:9}
\end{align}
where $\sigma_{\alpha}$ represents the $\sigma$ gate ($\sigma=I,X,Y$
and $Z$) on the $\alpha$-th qubit. With this encoding, the Hamiltonian
of the cluster $H_{c}\left(\{c_{i}\}\right)$ is decomposed by 
\begin{align}
H_{c} & =\sum_{i}\xi_{i}P_{i},\label{eq:12}
\end{align}
where the coefficient $\xi_{i}$ is a real number and $P_{i}=\sigma_{0}\otimes\sigma_{1}\otimes...$
is the tensor product of Pauli matrices. Hence the energy of the system
is parameterized as $E(\boldsymbol{\theta})=\sum_{i}\xi_{i}\langle P_{i}\rangle_{\theta}$,
where $\langle P_{i}\rangle_{\theta}=\langle g(\boldsymbol{\theta})|P_{i}|g(\boldsymbol{\theta})\rangle.$
The ground state can be ascertained by minimizing $E(\boldsymbol{\theta})$.
The optimal point, $\boldsymbol{\theta}_{0}$, yields the minimal
value $E(\boldsymbol{\theta}_{0})$. Consequently, the approximated
ground state is represented as $|\tilde{g}\rangle=U_{G}(\boldsymbol{\theta}=\boldsymbol{\theta}_{0})|\boldsymbol{0}\rangle$. 

\subsection{\protect\label{SecGF}Green's function}

The Green's function can be calculated in both time and frequency
domains. In the time domain, one can compute the real-time retarded
Green's function \cite{Gomes2023,GF_1st,GF_vt} or imaginary-time
Green's function \cite{GF_tau}. To obtain the spectral function,
these methods require additional processes, such as Fourier transform
or numerical analytic continuation. Based on Lehmann representation
\cite{GF_EOM}, the Green's function can be calculated directly in
the frequency domain using the variational quantum eigensolver \cite{SSVQE}
to find excitation state energies. In this work, we calculate the
real-time retarded Green's function, as we believe simulating a time-evolution
problem on a quantum circuit is more natural.

\begin{figure}
\includegraphics{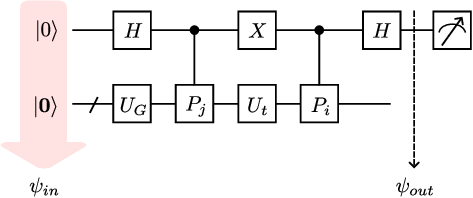}\caption{\protect\label{fig:circuitF} Following the fundamental principles
of the Hadamard test \cite{Htest}, the circuit is devised to compute
$\mathcal{F}\left(P_{i},P_{j}\right)$ as outlined in Eq. \ref{eq:14}.
The topmost qubit serves as an ancilla qubit and is configured for
the final measurement. The result of this measurement, denoted as
$p_{+}$ , determines the value of Eq. \ref{eq:14}.}
\end{figure}

The calculation of a retarded Green's function, based on the given
ground state $|g\rangle$, is defined by Eq. \ref{eq:4}. In the previous
calculation, additional parameter controls were needed to consider
the real and imaginary parts of each component of the Green's function
separately \cite{GF_1st}. In this context, we have restructured the
formulas so that only the computation of real values is necessary.
Using JW encoding, Eq. \ref{eq:4} is reformulated as:
\begin{align}
G_{ij}(t) & =\frac{(-1)^{i+j}}{4}\left(\mathcal{R}_{ij}-i\mathcal{I}_{ij}\right),\label{eq:11}
\end{align}
where 
\begin{align}
\mathcal{R}_{ij} & =\mathcal{F}(\bar{X_{i}},\bar{Y_{j}})-\mathcal{F}(\bar{Y_{i}},\bar{X_{j}})
\end{align}
 and 
\begin{align}
\mathcal{I}_{ij} & =\mathcal{F}(\bar{X_{i}},\bar{X_{j}})+\mathcal{F}(\bar{Y_{i}},\bar{Y_{j}}),
\end{align}
 represent the real and imaginary parts, respectively. In above expression,
we have defined $\bar{\sigma_{i}}=\prod_{\alpha<i}Z_{\alpha}\sigma_{i}$.
And $\mathcal{F}(\bar{\sigma_{i}},\bar{\sigma_{j}})$ is given by
\begin{align}
\mathcal{F}\left(\bar{\sigma_{i}},\bar{\sigma_{j}}\right)= & \langle g|e^{iH_{c}t}\bar{\sigma_{i}}e^{-iH_{c}t}\bar{\sigma_{j}}|g\rangle\nonumber \\
 & +\langle g|\bar{\sigma_{j}}e^{iH_{c}t}\bar{\sigma_{i}}e^{-iH_{c}t}|g\rangle.\label{eq:14}
\end{align}
The advantage of basing all calculations on Eq. \ref{eq:14} is that
$\mathcal{F}\left(\bar{\sigma_{i}},\bar{\sigma_{j}}\right)$ is a
real number, and Eq. \ref{eq:14} can be directly computed using the
circuit shown in Fig. \ref{fig:circuitF}. This circuit, which includes
parameters $i,j,$$t$ and $N_{\tau}$, is repeatedly called as a
subroutine by the program. At the end, the attached qubit on the top
line in Fig. \ref{fig:circuitF} is measured. The probability of measuring
0, denoted as $p_{+}$, is obtained through sampling from the quantum
circuit. The final value of Eq. \ref{eq:14}, is given by $\mathcal{F}\left(\bar{\sigma_{i}},\bar{\sigma_{j}}\right)=2(2p_{+}-1)$,
with a detailed derivation available in Appendix \ref{Apd_F}.

In this approach, to simulate the time-evolution of $U_{t}=e^{-iH_{c}t}$,
usually Suzuki-trotter \cite{Suzuki} decomposition is used. It is
given by
\begin{align}
U_{t} & =\left(e^{-iH_{c}\frac{t}{N_{\tau}}}\right)^{N_{\tau}}\approx\left(\prod_{j}e^{-i\xi_{j}P_{j}\frac{t}{N_{\tau}}}\right)^{N_{\tau}}.\label{eq:19}
\end{align}

To achieve the necessary precision, it is imperative to use rather
small time slices $\tau=t/N_{\tau}$, which results in a very long
circuit depth. However, the current state of quantum hardware, with
its noise levels, poses a challenge in simulating such long circuits
with the required accuracy. A potential workaround is to employ variational
principles, such as McLachlan's \cite{McLachlan_1964,PhysRevX.7.021050},
to approximate the original decomposition circuit with a shorter one
\cite{GF_1st,GF_vt}. Nevertheless, this introduces computational
overhead, as fitting an ansatz circuit to replicate the time evolution
of the Hamiltonian requires solving an additional system of differential
equations \cite{PhysRevX.7.021050}. These challenges raise concerns
about the scalability of this method as the system size increases.
In this work, to determine the noise threshold necessary for the direct
application of the Suzuki decomposition, we persist in employing Eq.
\ref{eq:19} to calculate the Green's function on a noisy simulator
with a comprehensive depolarization error \cite{Nielsen2010} rate
of 0.01\%, affecting each gate and channel. To maintain consistency
in the subsequent calculations, we utilize an exact ground state $|g\rangle$
as input, despite having previously obtained a ground state $|\tilde{g}\rangle$
through VQE on quantum hardware.

\section{\protect\label{SecSteps} Computation procedures}

All of our numerical calculations are based on isQ \cite{10129229}.
In the preceding sections, we delved into methodologies for determining
the ground state and computing the Green's function. We summarize
below the standard steps for the whole calculation:
\begin{enumerate}
\item \textbf{Initialization of the ansatz}: Set the ansatz $U_{G}(\boldsymbol{\theta})$,
and based on the variational principle by calling quantum hardware
iteratively, a set of optimized parameters $\boldsymbol{\theta}_{0}$
(therefore the optimized ground state $|\tilde{g}\rangle=U_{G}(\boldsymbol{\theta}_{0})|\boldsymbol{0}\rangle$)
is found. 
\item \textbf{Calculation of the retarded Green's function}: For the specified
input $|g\rangle$, we employ the circuit illustrated in Fig. \ref{fig:circuitF}
to compute Eq. \ref{eq:14}. Subsequently, this allows us to calculate
the real-time retarded Green's function as defined in Eq. \ref{eq:11}.
\item \textbf{Fourier transformation for frequency domain}: The cluster
Green's function in the frequency domain, $G_{ij}(\omega)$, is derived
via a Fourier transform, as outlined in Eq. \ref{eq:10}. We adopt
the Gauss--Legendre quadrature method \cite{Numerical77} for implementing
an exceptionally precise and efficient integration.
\item \textbf{Computation of the lattice Green's function}: Following the
acquisition of each matrix element $G_{ij}$ of $\boldsymbol{G}$,
proceed to calculate the lattice Green's function $\boldsymbol{\mathcal{G}}$
by applying Eq. \ref{eq:3}.
\end{enumerate}

\section{Results and discussions}

\subsection{\protect\label{SecGSE} Ground state energy }

\begin{figure}
\includegraphics[scale=0.43]{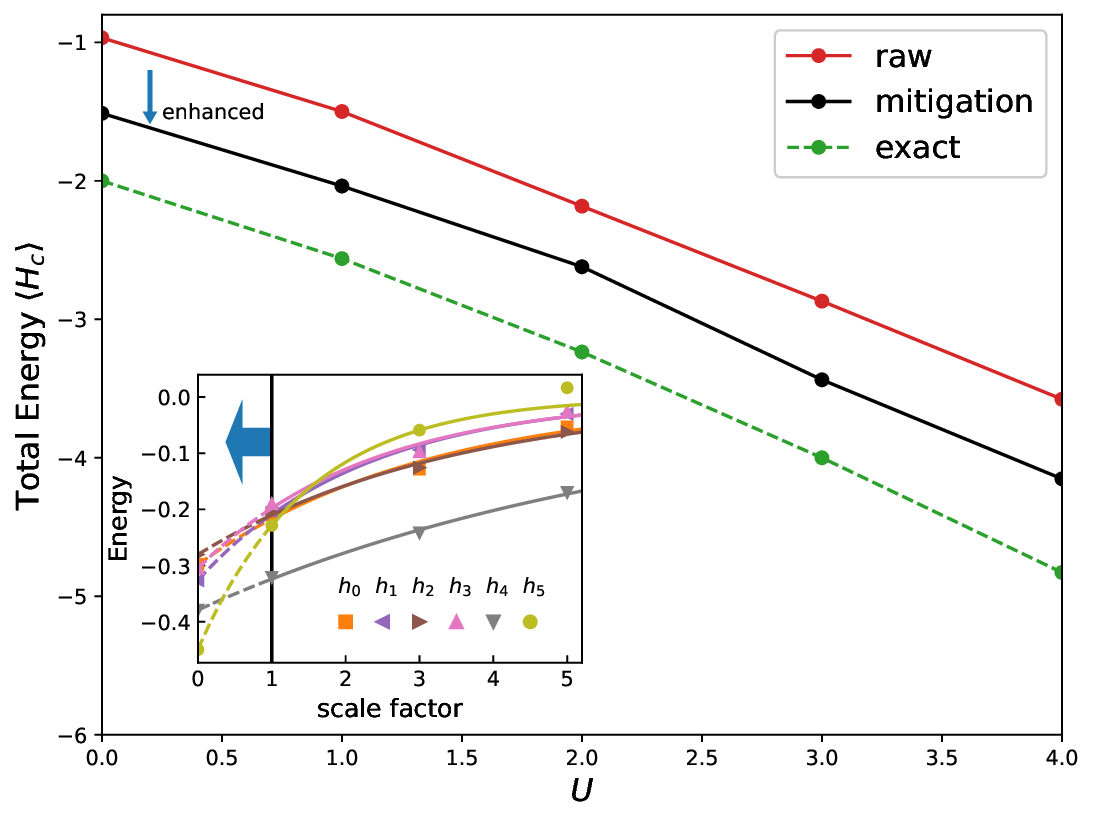}\caption{\protect\label{fig:energy} The ground state energies as a function
of interaction $U$. Inset details illustrate the mitigation within
Hamiltonian $H_{c}$ at $U=3$, showing the contribution of each term.}
\end{figure}

For clarity and simplicity, we omit the consideration of symmetry-breaking
cases, thereby reverting the calculation to a standard CPT calculation.
Meanwhile, since the reference system lacks bath sites, our study
is limited to the insulator phase. Despite these simplifications,
we still possess sufficient computational resources to consider the
anti-ferromagnetic phase \cite{AF}, having selected a $2\times1$
cluster as the reference system, as illustrated in Fig. \ref{fig:qcm}.
When we consider introducing variational parameters to handle symmetry-breaking
scenarios, we only need to make repeated calls to our existing computational
process under varying parameters, incurring a slight increase in computational
effort. This approach does not fundamentally increase the overall
computational burden. 

With consideration of the spin degree of freedom, the quantum state
of the cluster is encoded by four qubits. In our calculations, we
set the hopping energy $\gamma=1$ as the energy unit, and $\mu=\frac{U}{2}$
to maintain the half-filling condition. In our case, in pursuit of
high efficiency, we employ a well-designed ansatz that consists solely
of a single variational parameter, denoted as $\varphi$ . The variational
process is further simplified by directly minimizing analytical sinusoidal
function which is obtained by fitting sparse sampling data from scanning
$\varphi$ \cite{fit1}. More details of the variational ansatz can
be found in Appendix\textcolor{red}{{} \ref{Apd_Ansatz}.}

The final total energies of the cluster system as a function of $U$
are shown in Fig. \ref{fig:energy}. The raw data of energy (red solid
line), which is calculated from the original circuit on the quantum
hardware without mitigation, decreases with increasing U. The exact
result by classical simulation is calculated and shown as the green
dashed line in Fig. \ref{fig:energy}. We see that the raw data is
consistent with the exact result quantitatively despite some discrepancies.
We recorrect this result by post-condition of digital zero noise extrapolation
(DZNE) with circuit folding \cite{DZNE}. For the $U$>0 case, there
are 6 terms, in addition to a constant term with $P=I$, in the Hamiltonian,
see Eq. \ref{eq:12}. The result of each term at $U=3$ are shown
in the insertion of Fig. \ref{fig:energy}, from which we see that
these results have been improved to varying degrees. The corrected
energy is shown as the black solid line in Fig. \ref{fig:energy}.
Compared with the raw data, the result is enhanced significantly.

\subsection{\protect\label{SecCGF}Green's function and spectral function}

\begin{figure}
\includegraphics[scale=0.53]{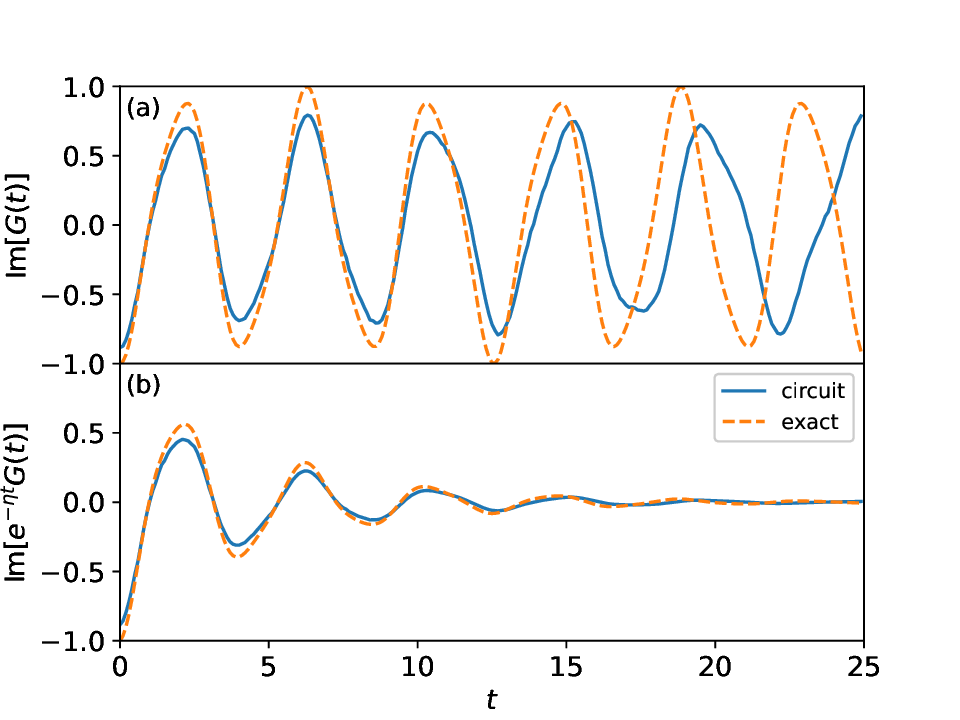}\caption{\protect\label{fig:gt} Comparison of retarded Green's function calculated
by a quantum circuit (cyan solid line) and the exact result (orange
dashed line). (a) The imaginary part of the real-time retarded Green's
function. (b) The Green's function after being multiplied by a factor
$e^{-\eta t}$ which effectively reduces the error at large $t$.}
\end{figure}

In our subsequent analyses, we maintain $U=3$. The computation of
the real-time retarded Green's function necessitates a delicate balance
in the Suzuki approximation's precision. Specifically, the time slice
$\tau$ in the decomposition of $e^{-iHt}$ (refer to Eq. \ref{eq:19})
must be carefully calibrated. If $\tau$ is excessively large, precision
is compromised, while if it is too small, the cumulative errors from
a circuit with long depth would overshadow any meaningful data obtained
from the results. To address this, we standardized $\tau$ with $N_{\tau}=60$
for all simulations.

As an illustrative case, we computed the on-site Green's function
at the first site $(i=j=1)$ for a single spin. Due to symmetry, the
results for spin up and spin down are the same. Given the particle-hole
symmetry in the system, the real part of this Green's function vanishes,
and we have exclusively calculated and presented the imaginary part,
$\text{Im}[G(t)]$, represented by the solid cyan lines in Fig. \ref{fig:gt}.
For comparative analysis, we also calculated the exact result, represented
as orange dashed lines in the same figure. As shown in Fig. \ref{fig:gt}
(a), in scenarios with small $t$, the circuit-derived results closely
align with the exact results, maintaining satisfactory precision.
This alignment is attributed to the smaller value of $\tau$ (due
to the fixed value of $N_{\tau}$), ensuring the accuracy of the Suzuki
decomposition. Minor discrepancies in the small $t$ region primarily
stem from circuit noise. However, as $t$ increases, $\tau$ consequently
becomes larger, leading to a decline in the accuracy of the Suzuki
approximation. This trend is observable in Fig. \ref{fig:gt} (a),
where beyond $t\approx22$, the circuit results significantly diverge
from the exact ones. Despite this, we demonstrate that such large
discrepancies at higher $t$ values do not substantially affect subsequent
calculations. Recall Eq. \ref{eq:10}, where the integrand includes
the factor $e^{-\eta t}$. For practical computations, we have set
$\eta=0.2$. The exponential decay induced by $e^{-\eta t}$ effectively
mitigates the impact of discrepancies resulting from larger $\tau$
values at higher $t$ regions, as illustrated in Fig. \ref{fig:gt}
(b). This property can serve as a valuable guide for devising more
refined computational strategies, allowing us to balance precision
with the allocation of computational resources effectively.

\begin{figure}
\includegraphics[scale=0.53]{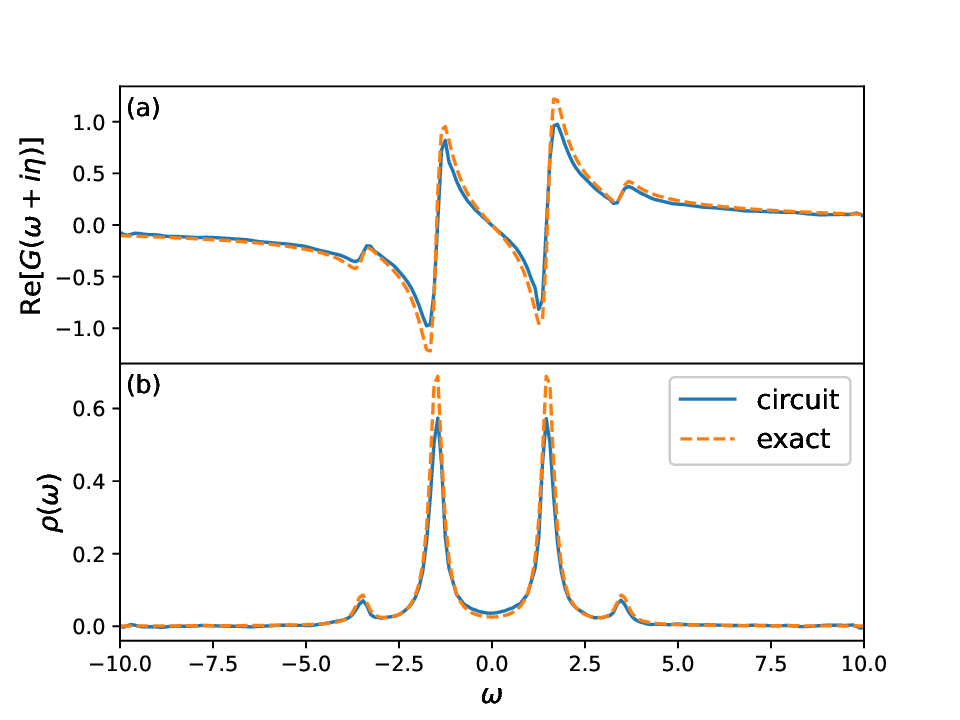}\caption{\protect\label{fig:gw} The quantum circuit calculations (cyan solid
line) and exact results (orange dashed line). (a) The real part of
the Green's function in the frequency domain. (b) The corresponding
spectral function of the cluster.}
\end{figure}

To perform the integration specified in Eq. \ref{eq:10}, we employ
a total of $100$ sampling points and set the upper limit to $t_{\infty}=30$.
We have ascertained the sufficiency of this integration range through
our prior discussion, as illustrated in Fig. \ref{fig:gt} (b). The
real part of $G(\omega+i\eta)$ and the corresponding spectral function
are presented in Fig. \ref{fig:gw} (a) and (b), respectively. We
depict the results obtained through the circuit model as solid cyan
lines and compare them with the exact results represented by dashed
orange lines. It is noteworthy that, despite the presence of notable
discrepancies arising from circuit noise in the calculation of the
real-time retarded Green's function, the ultimate outcome---$G$
in the frequency domain---is within reasonable accuracy. The positions
of the poles of the Green's function remain accurate, albeit with
a decrease in amplitudes attributable to the effects of noisy errors.
Our examination of the summation rule reveals that $\int d\omega\rho(\omega)\sim0.88$,
indicating an approximate $12\%$ loss. This diminution is also visually
evident in Fig. \ref{fig:gw} (b), where the inner peaks appear marginally
reduced. This loss predominantly stems from quantum noise.

\subsection{\protect\label{SecAK}One-particle excitation spectra}

\begin{figure}
\includegraphics[scale=0.55]{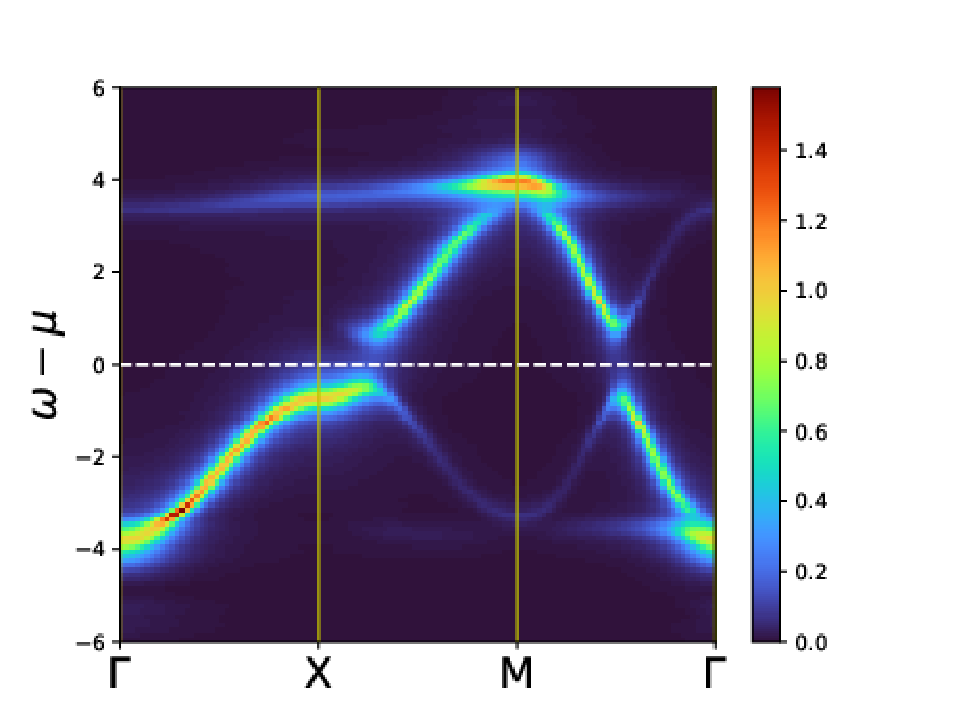}\caption{\protect\label{fig:ak} Intensity plot of one-particle excitation
spectra for the lattice system revealing the distribution and intensity
of excitations across various energy levels and momentum states.}
\end{figure}

Once we obtain the cluster Green's function, we can proceed with the
calculation of physical properties relevant to the lattice model.
The lattice Green\textquoteright s function in the original Brillouin
zone is defined as 
\begin{align}
g_{\sigma}(\boldsymbol{k},\omega) & =\frac{1}{L}\sum_{ij}e^{-i\boldsymbol{k}\cdot(\boldsymbol{r}_{i}-\boldsymbol{r}_{j})}\mathcal{G}_{i\sigma,j\sigma}(\boldsymbol{k},\omega),
\end{align}
where $\sigma=\uparrow$ or $\downarrow$ represents the spin index,
and $i,j$ span all sites within the cluster. The definition of the
single-particle excitation spectra $\rho_{\sigma}(\boldsymbol{k},\omega)$
remains the same as Eq. \ref{eq:2}. Due to the symmetry, we calculate
$\rho_{\uparrow}(\boldsymbol{k},\omega)$, and the results are shown
in Fig. \ref{fig:ak}. The obtained data captures information around
the typical critical points, $\Gamma$, X, and M, which are consistent
with the classical simulation of a Hubbard model on a 2D square lattice
by conventional methods. A key characteristic of this model, the Mott-gap,
is evident around the Fermion surface due to interaction $U$, as
depicted in Fig. \ref{fig:ak}. Despite some loss in spectral function
weight, clear band structures are still observable. This analysis
of one-particle excitation spectra paves the way for computing other
physical quantities, such as optical conductivity \cite{VCA2}. Such
calculations also hold significant practical relevance in experimental
contexts, as this spectra can be extracted through techniques like
angle-resolved photoemission spectroscopy (ARPES).

\section{\protect\label{SecConc} Conclusion }

In summary, we have thoroughly demonstrated the feasibility of studying
strongly correlated systems using quantum circuits. To account for
local correlations using the QCM, we segment the lattice model into
reference clusters to abstract self-energy. Employing the VQE with
an appropriate ansatz implemented on quantum hardware, we successfully
obtained the ground state of a reference cluster, further refining
the results through the error mitigation of DZNE method. Formulas
for calculating Green's functions within the circuit model were derived.
We observed that the real-time retarded Green's function results are
consistent with exact results within an acceptable error in the small
$t$ region, and the influence of any discrepancies in the large $t$
region on calculating the lattice Green's function has been suppressed.
Utilizing the lattice Green's function, we investigated the one-particle
excitation spectra. Our computations successfully reproduced the typical
properties of the band structure of a Hubbard model on a 2D square
lattice. 

\section{Author infomation}

\subsection*{Corresponding authors}

\paragraph*{Hengyue Li}

Arclight Quantum Computing Inc., Beijing, 100191, China; Email: \href{http://lihy@arclightquantum.com}{lihy@arclightquantum.com}

\paragraph*{Yusheng Yang}

Arclight Quantum Computing Inc., Beijing, 100191, China; Email: \href{http://yangys@arclightquantum.com}{yangys@arclightquantum.com}

\paragraph*{Shenggang Ying}

Arclight Quantum Computing Inc., Beijing, 100191, China; Institute
of Software, Chinese Academy of Sciences, Beijing, 100190, China;
Email: \href{http://yingsg@ios.ac.cn}{yingsg@ios.ac.cn}

\subsection*{Authors}

\paragraph*{Pin Lv}

China Telecom Quantum Information Technology Group Co., Ltd., Hefei,
230031, China

\paragraph*{Jinglong Qu}

High-Temperature Materials Institute, Central Iron and Steel Research
Institute, Beijing, 100081, China; Beijing GAONA Materials \& Technology
Co., Ltd., Beijing, 100081, China; Sichuan CISRI-Gaona Forging Co.,
Ltd., Deyang, Sichuan, 618099, China

\paragraph*{Zhe-Hui Wang}

QuantumCTek (Shanghai) Co., Ltd., Shanghai, 200120, China

\paragraph*{Jian Sun}

Arclight Quantum Computing Inc., Beijing, 100191, China

\section{Acknowledgments }

We express our sincere gratitude to the quantum cloud platform of
\textit{QuantumCTek Co., Ltd.} (\href{https://quantumctek-cloud.com}{https://quantumctek-cloud.com})
for their invaluable support in providing quantum computing resources,
which were instrumental in conducting the computational experiments
for this study. Without access to their quantum hardware infrastructure,
this work would not have been feasible. Additionally, we extend our
heartfelt appreciation to Guolong Cui from \textit{Arclight Quantum
Computing Inc.} for his guidance and support throughout the research
process. This research is sponsored by Beijing Nova Program 20220484128.

\section*{Appendix: Details of QCM \protect\label{Apd_qcm}}

Utilizing the index transformation outlined in Eq. \ref{eq:2.12},
we can reformulate the original Hamiltonian (Eq. \ref{eq:2.11}) as
\begin{align}
H & =\sum_{\boldsymbol{R}\alpha\boldsymbol{R}^{\prime}\beta}\mathcal{T}_{\alpha,\beta}^{\boldsymbol{R}-\boldsymbol{R}'}\psi_{\boldsymbol{R}\alpha}^{\dagger}\psi_{\boldsymbol{R}^{\prime}\beta}\nonumber \\
 & +\sum_{\boldsymbol{R}}\sum_{\alpha\beta\gamma\delta}X_{\boldsymbol{R}\alpha,\boldsymbol{R}\beta,\boldsymbol{R}\gamma,\boldsymbol{R}\delta}\psi_{\boldsymbol{R}\alpha}^{\dagger}\psi_{\boldsymbol{R}\beta}^{\dagger}\psi_{\boldsymbol{R}\gamma}\psi_{\boldsymbol{R}\delta}.\label{eq:A11}
\end{align}
In Eq. \ref{eq:A11}, $\mathcal{T}^{\boldsymbol{R}-\boldsymbol{R}'}$
is defined as a $L\times L$ matrix, where $L$ denotes the site of
the cluster. We can write $t_{\boldsymbol{R}\alpha,\boldsymbol{R}'\beta}=\mathcal{T}_{\alpha,\beta}^{\boldsymbol{R},\boldsymbol{R}'}$
as $\mathcal{T}_{\alpha,\beta}^{\boldsymbol{R}-\boldsymbol{R}'}$
because of the translation symmetry. We assume all interactions to
be local within clusters. If inter-cluster interactions exist, clusters
need to be decoupled using a mean-field approximation. The inter-site
Coulomb interaction $Vn_{i}n_{j}$, for instance, can be approximated
at the Hartree-Fock level as : $Vn_{i}n_{j}\approx V(n_{i}\langle n_{j}\rangle+\langle n_{i}\rangle n_{j}-\langle n_{i}\rangle\langle n_{j}\rangle)$.
The fluctuation term $(n_{i}-\langle n_{i}\rangle)(n_{j}-\langle n_{j}\rangle)$
is neglected in this approximation, introducing additional mean fields
$\{\langle n_{i}\rangle\}$ into the expression. These mean fields
can be determined either through iterative calculation or by minimizing
the self-energy functional $\Omega_{t}[\Sigma]$ \cite{VCA0,VCA2,QCM1}.
Translation symmetry implies that the Hamiltonian parameters for each
cluster $\boldsymbol{R}$ are identical, leading to $X_{\boldsymbol{R}\alpha,\boldsymbol{R}\beta,\boldsymbol{R}\gamma,\boldsymbol{R}\delta}=\chi_{\alpha\beta\gamma\delta}$. 

Further, the first term in Eq. \ref{eq:A11} is divided into inter-cluster
and intra-cluster parts. By incorporating the identity $1=\delta_{\boldsymbol{R}\boldsymbol{R'}}+(1-\delta_{\boldsymbol{R}\boldsymbol{R'}})$
into the first term of Eq. \ref{eq:A11}, we obtain: $\sum_{\boldsymbol{R}}\sum_{\alpha\beta}\psi_{\boldsymbol{R}\alpha}^{\dagger}\mathcal{T}_{\alpha,\beta}^{0}\psi_{\boldsymbol{R}\beta}+\sum_{\boldsymbol{R}\neq\boldsymbol{R}^{\prime}}\sum_{\alpha\beta}\psi_{\boldsymbol{R}\alpha}^{\dagger}\mathcal{T}_{\alpha,\beta}^{\boldsymbol{R}-\boldsymbol{R}'}\psi_{\boldsymbol{R}^{\prime}\beta}$.
Therefore, the Hamiltonian (Eq. \ref{eq:A11}) is decoupled by clusters
and can be rewritten as:
\begin{align}
H & =\sum_{\boldsymbol{R}}H_{c}\left(\{\psi_{\boldsymbol{R}\alpha}\}\right)+\sum_{\boldsymbol{R}\boldsymbol{R}'}T(\boldsymbol{R}-\boldsymbol{R}'),\label{eq:A12-1}
\end{align}
where 
\begin{align}
H_{c}\left(\{\psi_{\boldsymbol{R}\alpha}\}\right) & =\sum_{\alpha\beta}\psi_{\boldsymbol{R}\alpha}^{\dagger}\mathcal{T}_{\alpha,\beta}^{0}\psi_{\boldsymbol{R}\beta}\nonumber \\
 & +\sum_{\alpha\beta\gamma\delta}\chi_{\alpha\beta\gamma\delta}\psi_{\boldsymbol{R}\alpha}^{\dagger}\psi_{\boldsymbol{R}\beta}^{\dagger}\psi_{\boldsymbol{R}\gamma}\psi_{\boldsymbol{R}\delta},\label{eq:13}
\end{align}
and 
\begin{align}
T(\boldsymbol{R}-\boldsymbol{R}') & =\sum_{\alpha\beta}\psi_{\boldsymbol{R}\alpha}^{\dagger}\mathcal{T}_{\alpha,\beta}^{\boldsymbol{R}-\boldsymbol{R}'}\psi_{\boldsymbol{R}^{\prime}\beta}.
\end{align}
The subscript ``c'' in $H_{c}$ signifies that it represents the
Hamiltonian corresponding to a local cluster. For the non-interacting
case where $\chi_{\alpha\beta\gamma\delta}=0$, the Green's function
of the lattice Hamiltonian is computed by performing the Fourier transform:
\begin{align}
\psi_{\boldsymbol{q}\alpha} & =\frac{1}{\sqrt{M}}\sum_{\boldsymbol{R}}\psi_{\boldsymbol{R}\alpha}e^{-i\boldsymbol{q}\cdot\boldsymbol{R}},\label{eq:A15}
\end{align}
where $M=\frac{N}{L}$ is the number of clusters in the super-lattice
($N$ being the total number of sites in the original lattice), and
the summation over $\boldsymbol{R}$ covers all super-lattice cells,
with $\boldsymbol{q}$ as the reciprocal vector in the reduced Brillouin
zone. Using the inverse transform of Eq. \ref{eq:A15}, the non-interacting
Hamiltonian is given by $H_{0}=\sum_{\boldsymbol{q}}\psi_{\boldsymbol{q}}^{\dagger}\left(\mathcal{T}^{0}+\tau_{\boldsymbol{q}}\right)\psi_{\boldsymbol{q}}$,
where $\psi_{\boldsymbol{q}}=\left(\begin{array}{c}
\psi_{\boldsymbol{q}1}\\
...\\
\psi_{\boldsymbol{q}L}
\end{array}\right),$ and $\tau_{\boldsymbol{q}}=\sum_{\boldsymbol{r}}'e^{-i\boldsymbol{q}\cdot\boldsymbol{r}}\mathcal{T}^{\boldsymbol{r}}$,
with the summation $\sum_{\boldsymbol{r}}'$ representing the summation
over all connected clusters. In the non-interacting case, the Green's
function $\boldsymbol{\mathcal{G}}_{0}(\boldsymbol{q},\omega)$ is
given by:
\begin{align}
\boldsymbol{\mathcal{G}}_{0}^{-1}(\boldsymbol{q},\omega) & =\omega-\mathcal{T}^{0}-\tau_{\boldsymbol{q}}.\label{eq:16-1}
\end{align}
For the interacting case, according to the CPT the Green's function
$\boldsymbol{\mathcal{G}}(\boldsymbol{q},\omega)$ is:
\begin{align}
\boldsymbol{\mathcal{G}}^{-1}(\boldsymbol{q},\omega) & =\boldsymbol{\mathcal{G}}_{0}^{-1}(\boldsymbol{q},\omega)-\boldsymbol{\Sigma}(\omega),\label{eq:17-1}
\end{align}
where $\boldsymbol{\Sigma}(\omega)$ represents the self-energy abstracted
from a reference system $H_{c}\left(\{f_{i}\}\right)$. In cases of
symmetry breaking, mean fields $\Delta$ are added to the reference
system \cite{VCA0}. The reference system's Green's function for the
interacting case is therefore given as:
\begin{align}
\boldsymbol{G}^{-1}(\omega) & =\omega-\mathcal{T}^{0}-\boldsymbol{\Sigma}(\omega).\label{eq:18-1}
\end{align}
Following Eqs. \ref{eq:16-1}\ref{eq:17-1}, and \ref{eq:18-1}, we
arrive at the relation:
\begin{align}
\boldsymbol{\mathcal{G}}^{-1}(\boldsymbol{q},\omega) & =\boldsymbol{G}^{-1}(\omega)-\tau_{\boldsymbol{q}}.
\end{align}

\section*{Appendix: Circuit for $\mathcal{F}\left(P_{i},P_{j}\right)$\protect\label{Apd_F}}

In Figure \ref{fig:circuitF}, the quantum circuit is initialized
with the input state $|\psi_{in}\rangle=|0\rangle|\boldsymbol{0}\rangle$.
The output state, before the final measurement, is represented as:
\begin{align}
|\psi_{out}\rangle & =\frac{1}{2}|0\rangle\left(P_{i}U_{t}+U_{t}P_{j}\right)|g\rangle\nonumber \\
 & +\frac{1}{2}|1\rangle\left(U_{t}P_{j}-P_{i}U_{t}\right)|g\rangle.
\end{align}
When measuring the ancilla qubit, the probabilities of observing outcomes
0 and 1 are denoted as $p_{+}$ and $p_{-}$, respectively. These
probabilities are calculated as follows:
\begin{align}
p_{\pm} & =\frac{1}{4}\langle g|\left[2\pm\left(P_{j}U_{t}^{\dagger}P_{i}U_{t}+U_{t}^{\dagger}P_{i}U_{t}P_{j}\right)\right]|g\rangle.
\end{align}
Consequently, the difference between $p_{+}$ and $p_{-}$ is given
by:
\begin{align}
p_{+}-p_{-} & =\frac{1}{2}\langle g|\left(P_{j}U_{t}^{\dagger}P_{i}U_{t}+U_{t}^{\dagger}P_{i}U_{t}P_{j}\right)|g\rangle,
\end{align}
This leads to the derivation of $\mathcal{F}\left(P_{i},P_{j}\right)=2(p_{+}-p_{-})\xrightarrow{p_{+}+p_{-}=1}2(2p_{+}-1).$

\section*{Appendix: Ansatz \protect\label{Apd_Ansatz}}

\begin{figure}
\includegraphics[scale=0.8]{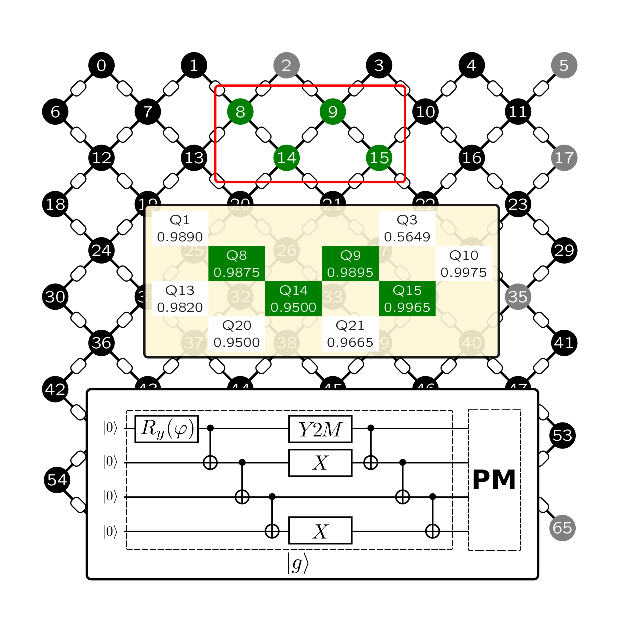}\caption{\protect\label{figcir2} The quantum circuit layout employed in our
experiment. Selected qubits are interconnected and notable for their
high readout fidelity, as highlighted in the middle section of the
diagram. The circuit features a custom-designed ansatz, adapted specifically
for our search problem. PM= Pauli Measurement and $Y2M=R_{y}(-\frac{\pi}{2})$
is a dedicated gate in \textit{Xiaohong. }}
\end{figure}

In our experiments, we configured the circuit sampling to consist
of 12,000 shots. Figure \ref{figcir2} presents the qubits selected
for our study. These qubits, interconnected with adjacent ones, exhibited
outstanding quality during the experiment, as indicated by the high
readout fidelity depicted in the central part of the figure. We developed
a specialized ansatz tailored to our search problem, as shown in Fig.
\ref{figcir2}. Given the symmetry inherent in our problem, we incorporated
a single variable parameter, implemented using an $R_{y}$ gate. To
generate the necessary entanglement, two arrays of CNOT gates were
employed. Our design permits experimentation with various single-qubit
gate configurations interspersed between these CNOT gate arrays. For
instance, substituting the $X$ gates with $R_{y}(\pi)$ and $R_{y}(-\pi)$,
we derived another effective ansatz. We believe that a more simplified
ansatz is achievable. 

\section*{Recorrection log}

\subsection*{2025-12-23}

This revised preprint corrects errors in Equation (14) and Equation
(15) of the paper published in \textit{Phys. Scr. 99 105117 (2024)}.
These two equations were incorrectly formulated in the original publication.
It is important to clarify that these errors do not hinder the overall
understanding of the paper\textquoteright s content, nor do they affect
the accuracy of any computational results and core conclusions presented
in the work. 

\bibliographystyle{unsrt}
\bibliography{reference}

\end{document}